\documentclass[aps,twocolumn]{revtex4}
\usepackage[dvips]{graphics,graphicx}
\usepackage{amsmath}

\usepackage{amssymb}
\usepackage{epstopdf}
\usepackage[usenames, dvipsnames]{color}
\usepackage{graphicx}
\usepackage{graphics}
\usepackage{amsfonts}
\usepackage{eufrak}
\usepackage{mathrsfs}
\usepackage{bm}

\begin{document}

\title{Resonant transport of bosonic carriers through a quantum device}
\author{ P. S. Muraev$^{1,2}$, D. N. Maksimov$^{1,3}$, and A. R. Kolovsky$^{1,2}$}
\affiliation{$^1$Kirensky Institute of Physics, Federal Research Center KSC SB
RAS, 660036, Krasnoyarsk, Russia}
\affiliation{$^2$School of Engineering Physics and Radio Electronics,
Siberian Federal University, 660041, Krasnoyarsk, Russia}
\affiliation{$^3$IRS SQC, Siberian Federal University, 660041, Krasnoyarsk, Russia}

\date{\today}
\begin{abstract}
We analyze the current of Bose particles across the tight-binding chain connected at both ends to the particles reservoirs. Unlike the standard open Bose-Hubbard model, where the presence of reservoirs is taken into account by the Lindbladians  acting on the first and the last sites of the chain, we use the semi-microscopic models for the reservoirs. This allows us to address the case of arbitrary reservoir temperature.  In particular, we discuss the phenomenon of the resonant transmission for nearly condensed bosons, where the current  across the chain is significantly enhanced for certain values of the gate voltage.
\end{abstract}
\maketitle

\section{Intoduction}
\label{sec1}

Recently we have witnessed the increase of interest to the open (dissipative) Bose-Hubbard (BH) system which has become the paradigm model for quantum transport with Bose particles \cite{Baro13,Ivan13,Kord15a,Kord15b,Haco15,Labo15,Labo16,112,Fedo21}. Experimentally, there are two main platforms for realizing the open BH model: the superconducting circuits, namely, the chain of coupled transmon qubits \cite{Haco15,Fedo21} and the cold Bose atoms in optical lattices \cite{Baro13,Labo15,Labo16}. For studying the quantum transport in the former system photons are injected into the first transmon of the chain by using a microwave generator and the signal is read from the last transmon in the chain. In the latter system one measures the atomic current across the lattice connecting two atomic reservoirs with different chemical potentials, in spirit of the laboratory experiment  \cite{Lebr18} conducted with the Fermi atoms.  The unique  feature of the bosonic system, however, is that with the increase of the carriers density it shows a transition from quantum to classical regime where the BH chain can be viewed as system of coupled classical oscillators \cite{116}. This quantum-to-classical transition for the identical particles motivates further studying transport phenomena with bosonic carriers. 

In our previous works \cite{112,116} we analyzed the current of Bose particles within the framework of the standard open BH model where the effect  of reservoirs is taken into account by introducing the gain and loss Lindbladian operators acting on the first and the last sites of the  chain. In the classical approach this corresponds to the situation where the first and the last oscillators are subject to the friction and are excited by white noise whose intensity is determined by the mean particle density of the respective reservoir \cite{116}. Unfortunately, the standard open BH model implies validity of the Markov approximation which is not justified for low-temperature reservoirs with nearly condensed  Bose particles. In the present work we overcome this problem by introducing a non-Markovian open BH model which is the bosonic analogue of the non-Markovian open Fermi-Hubbard model discussed in Ref.~\cite{120}.  It is shown below that in the pseudoclassical approach the proposed model instead of  white noise utilizes a narrow-band noise with the well defined mean frequency. This brings us closer to the experiment \cite{Fedo21} and, simultaneously,  to the situation one meets in the solid-state physics for fermions with well defined Fermi energy. In particular, similar to the fermionic case, we can address the phenomenon of the resonant transmission \cite{Datt95}.

\section{Non-Markovian open Bose-Hubbard model}
\label{sec2}

Since we are interested in the non-Markovian case a microscopic model for the particle reservoirs is required. Following Ref.~\cite{120} we use as the reservoirs the tight-binding rings of size $M$ each, where eventually $M\rightarrow\infty$. These rings are attached to both ends of the BH chain of length $L$.  Bosons can hop between the sites of the chain and the sites of the rings with the rates $J_{\rm s}$  and $J_{\rm r}$, where $J_{\rm s}\sim J_{\rm r}$, while the hopping between the chain and the rings is controlled  by the coupling constant $\epsilon\ll J_{\rm s}, J_{\rm r}$. If the chemical potential or temperature of the left  and right rings are different we have a directed current across the chain. 
\begin{figure}[b]
\includegraphics[width=7.5cm,clip]{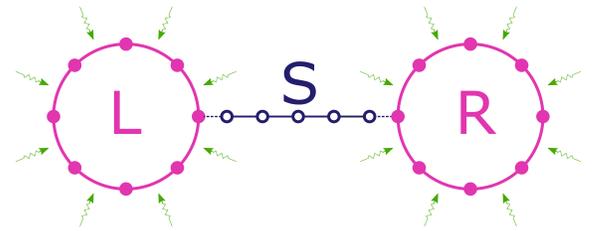}
\caption{Schimetic presentation of the model.}
\label{fig0}
\end{figure}

\subsection{Quantum Hamiltonians and the governing equation}

The system dynamics is governed by the master equation for the total density matrix
\begin{equation}
\label{a1}
\frac{\partial \widehat{{\cal R}}}{\partial t}=-i[\widehat{{\cal H}},  \widehat{{\cal R}}]+
\sum_{j={\rm L},{\rm R}}\left[\widehat{{\cal L}}_g^{(j)}(\widehat{{\cal R}}) +\widehat{{\cal L}}_d^{(j)}(\widehat{{\cal R}}) \right] \;,
\end{equation}
where  the Hamiltonian $\widehat{{\cal H}}$ has the form
\begin{equation}
\label{a2}
\widehat{{\cal H}}=\widehat{{\cal H}}_{\rm s}
+ \sum_{j={\rm L},{\rm R}} \widehat{{\cal H}}_{\rm r}^{(j)} +   \sum_{j={\rm L},{\rm R}} \widehat{{\cal H}}_\epsilon^{(j)} \;.
\end{equation}
For the chain Hamiltonian $\widehat{{\cal H}}_{\rm s}$ we have
\begin{equation}
\label{a3}
\widehat{{\cal H}}_{\rm s}=\delta \sum_{\ell=1}^{L}\hat{n}_{\ell}
-\frac{J_{\rm s}}{2}\left(\sum_{\ell=1}^{L-1}\hat{a}_{\ell+1}^{\dagger}\hat{a}_{\ell} +{\rm h.c.} \right)
+\frac{U}{2}\sum_{\ell=1}^{L}\hat{n}_{\ell} (\hat{n}_{\ell}-1)
\end{equation}
with $\hat{a}_{\ell}^{\dagger},\hat{a}_{\ell}$ being bosonic creation and annihilation operators at the $\ell {\rm th}$ site. For future purpose we also included in the chain Hamiltonian the on-site energy (gate voltage) $\delta$ and the inter-particle interaction whose strength is characterised by the microscopic interaction constant $U$.  The ring Hamiltonians,  $\widehat{{\cal H}}_{\rm r}$,  and the coupling  Hamiltonians, $\widehat{{\cal H}}_\epsilon$, are indexed with superscript $j={\rm L},{\rm R}$ specifying the ring and we write the ring Hamiltonians in terms of bosonic operators acting in the Fock space of the Bloch states,
\begin{equation}
\label{a4}
\widehat{{\cal H}}_{\rm r}=
\sum_{k=1}^{M} E_k \hat{b}_{k}^{\dagger}\hat{b}_{k} \;, \quad E_k=-J_{\rm r}\cos\left(\frac{2\pi k}{M}\right) \;.
\end{equation}
Here and below we dropped superscript $j$ assuming that the rings are identical. The coupling Hamiltonian is given by
\begin{equation}
\label{a5}
\widehat{{\cal H}}_\epsilon=
-\frac{\epsilon}{2 \sqrt{M}}  \hat{a}_\ell^\dagger \sum_{k=1}^M\hat{b}_k +{\rm h.c.}
\end{equation}
where $\ell=1$ for $j={\rm L}$ and  $\ell=L$ for $j={\rm R}$. For the sake of simplicity we set in what follows $J_{\rm s}=J_{\rm r}\equiv J$ which, in turn, is set to unity. Thus all energy constants are measured in units of $J$. Also, if not stated otherwise,  $\delta=0$ and $U=0$.

To prescribe thermodynamic properties to the reservoirs we introduce the particle drain,
\begin{equation}
\label{a6}
\widehat{{\cal L}}_d(\widehat{{\cal R}})=-\frac{\gamma}{2}\sum_{k=1}^M (\bar{n}_k+1)
\left(\hat{b}_{k}^{\dagger}\hat{b}_{k}\widehat{\cal R }-2\hat{b}_{k}\widehat{\cal R }\hat{b}_{k}^{\dagger}
+\widehat{\cal R }\hat{b}_{k}^{\dagger}\hat{b}_{k} \right) \;,
\end{equation}
and the particle gain,
\begin{equation}
\label{a7}
\widehat{{\cal L}}_g(\widehat{{\cal R}})=-\frac{\gamma}{2}\sum_{k=1}^M \bar{n}_k
\left(\hat{b}_{k}\hat{b}_{k}^{\dagger}\widehat{\cal R }-2\hat{b}_{k}^{\dagger}\widehat{\cal R }\hat{b}_{k}
+\widehat{\cal R }\hat{b}_{k}\hat{b}_{k}^{\dagger} \right) \;,
\end{equation}
where 
\begin{equation}
\label{a8}
\bar{n}_k=\frac{1}{e^{\beta(E_k +\mu)} - 1}  \;.
\end{equation}
These Lindbladians ensure the occupation of the Bloch states of the isolated ($\epsilon=0$) ring relaxing to the Bose-Einstein distribution with given chemical potential $\mu$ and inverse temperature $\beta$.  The rate at which this  relaxation takes place is determined by the constant $\gamma$. In what follows we use as the control parameter the particle density $\bar{n}=\sum_k \bar{n}_k/M$  which together with the temperature uniquely determines the chemical potential $\mu$. We denote the particle density in the left and right rings by $\bar{n}_{\rm L}$ and $\bar{n}_{\rm R}$, respectively.

Our main object of interest is the single-particle density matrix (SPDM) of bosons in the chain which is defined as 
\begin{equation}
\label{a9}
\rho_{\ell,m}(t)={\rm Tr}[\hat{a}_\ell^\dagger\hat{a}_m \widehat{{\cal R}}(t)] \;,\quad 1\le \ell,m \le L \;.
\end{equation}
Knowing the SPDM one finds the current in the chain (more precisely, the current density) by using the relation
\begin{equation}
\label{a10}
j(t)=\frac{1}{L-1}{\rm Tr}[\hat{\rho}(t)\hat{j}]  \;,
\end{equation}
where $\hat{j}$ is the current operator with the matrix elements $j_{\ell,m}=J_{\rm s}(\delta_{\ell,m+1}-\delta_{\ell,m-1})/2i$. In the case of non-interacting bosons ($U=0$) the dynamics of SPDM Eq.~(\ref{a9}) can be calculated by using, at least, two different methods. First, one can derive from the original master equation (\ref{a1}) the master equation for the total single-particle density matrix of the size $(M+L+M)\times(M+L+M)$   which is then  easily solved numerically. The central block of this matrix obviously  corresponds to the matrix (\ref{a9}). Below we use this method to calculate Fig.~\ref{fig2} and the dashed line in Fig.~\ref{fig4}.  The second method employs the pseudoclassical approximation to solve Eq.~(\ref{a1}).  The main advantage of the pseudoclassical approach  is that it can be equally applied to both non-interacting (where it is exact) and interacting bosons. Additionally, it provides a deeper insight into physics of the considered phenomena. We recollect the main points of the pseudoclassical approach in the next subsection.

\subsection{Pseudo-classical approach}

The pseudoclassical approach substitute the master equation  Eq.~(\ref{a1}) by the Fokker-Planck equation for the classical distribution function $f$ \cite{116}. Considering for the moment the case of a single ring (generalization onto the case of two ring is given in the beginning  of Sec.~\ref{sec3}),  the distribution function $f$ is the function of the time and $M$ canonical variables $b_k$ and $L$ canonical variables $a_\ell$ . The governing equation reads
\begin{equation}
\label{b1}
\frac{\partial f}{\partial t}=\{H,f\} + \sum_k\left[{\cal G}_k(f) + {\cal D}_k(f) \right] \;.
\end{equation}
In Eq.~(\ref{b1}) $\{\ldots,\ldots\}$ denotes the Poisson brackets, $H$ is the classical Hamiltonian of the system (which is obtained from the quantum Hamiltonian (\ref{a2}) by substituting the creation and annihilation operators with the classical canonical variables) and the last term is the Weyl symbol of the sum of the drain and gain Lindblad operators (\ref{a6}) and (\ref{a7}).  Explicitly we have 
\begin{equation}
\label{b4}
{\cal G}_k(f)=\frac{\gamma}{2}\left( b_k\frac{\partial f}{\partial b_k} + 2f + b_k^*\frac{\partial f}{\partial b_k^*} \right) \;,
\end{equation}
and  
\begin{equation}
\label{b5}
{\cal D}_k(f)= \gamma \left(\bar{n}+\frac{1}{2}\right)\frac{\partial^2 f}{\partial b_k \partial b_k^*}  \;.
\end{equation}
It is easy to show that Eq.~(\ref{b4}) corresponds to the friction (more precisely, contraction of the phase space volume) while Eq.~(\ref{b5}) describes the diffusion. Thus, one can put into correspondence to the Fokker-Planck equation (\ref{b1}) the following Langevin  equation
%
\begin{eqnarray}
   \label{c1}
i{\rm d} b_k=\left(E_k -i\frac{\gamma}{2}\right) b_k  {\rm d}t + \sqrt{\frac{\gamma \bar{n}_k}{2}} {\rm d}\xi_k +\frac{\epsilon}{2\sqrt{M}} a_1 {\rm d}t  \;,\\
   \label{c2}
i{\rm d}a_1=-\frac{J_{\rm s}}{2} a_2{\rm d}t + \frac{\epsilon}{2} {\rm d}\chi \;,\quad  \chi(t)=\frac{1}{\sqrt{M}} \sum_k b_k \;,\\
   \label{c3}
i{\rm d}a_\ell=-\frac{J_{\rm s}}{2}(a_{\ell-1}+a_{\ell+1}){\rm d}t \;,\quad \ell\ne 1 \;.
\end{eqnarray}
where $\xi_k$ are independent $\delta$-correlated random functions,   
$\langle {\rm d}\xi_k(t){\rm d}\xi_{k'}(t')\rangle=2\delta_{k,k'}\delta(t-t'){\rm d}t$.
\begin{figure}
\includegraphics[width=8.5cm,clip]{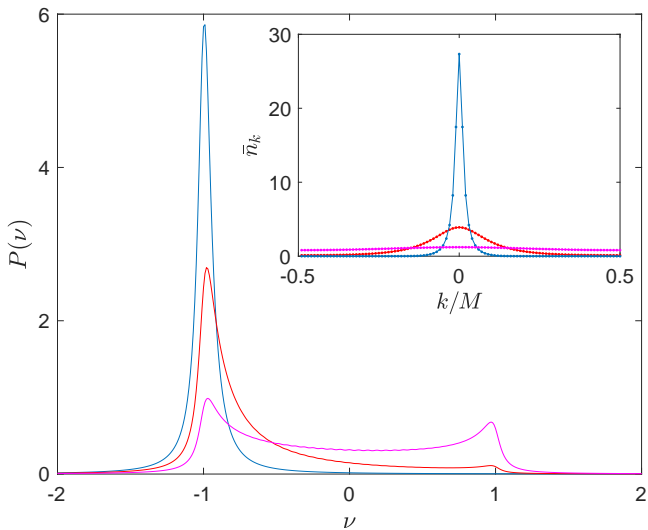}
\caption{The spectral density of the stochastic force $\chi(t)$ for $\bar{n}=1$ and $\beta=0.1,1,10$, from top to bottom at $\nu=0$. The inset shows the Bose-Einstein distribution for the same temperatures.}
\label{fig1}
\end{figure}

Let us discuss the displayed equations in more detail. For $\epsilon=0$ Eq.~(\ref{c1}) is  the damped harmonic oscillator subject to white noise. It has the steady-state solution where $\langle b_k^* b_k\rangle=\bar{n}_k$. Eqs.~(\ref{c2}-\ref{c3}) are the equations of motion for the chain of coupled linear oscillators where the first oscillator is subject to the stochastic force $\chi(t)$. We shall characterize this stochastic force by its spectral density $P(\nu)= |\chi(\nu)|^2$ where $\chi(\nu)$ is the Fourier transform of $\chi(t)$. The spectral density $P(\nu)$ is shown in Fig.~\ref{fig1} for $\gamma=0.1$ and three different values of the inverse temperature $\beta=0.1,1,10$. It is seen in Fig.~\ref{fig1} that condensation of bosons in the low-energy Bloch states results in the change of  $\chi(t)$ from a broad-band noise to a narrow-band noise. We also mention that the further decrease of the temperature below $1/\beta=0.1$ does not affect the displayed curve because in the limit $\beta\rightarrow\infty$ its shape is determined by the value of $\gamma$ but not by the width of the quasimomentum distribution (which tends to $\delta$-function). The same is true for  temperatures larger than $1/\beta=10$ because the quasimomentum distribution is practically flat already for $\beta=0.1$.

\section{Stationary current}
\label{sec3}

We proceed  with analysis of the stationary current $\bar{j}= j(t\rightarrow\infty)$. To address the transport problem in the framework of the pseudoclassical approach Eqs.~(\ref{c1}-\ref{c3}) should be complemented by the equation for the last site of the chain,
%
\begin{equation}
\label{c4}
i{\rm d}a_L=-\frac{J_{\rm s}}{2} a_{L-1} {\rm d}t+ \frac{\epsilon}{2} {\rm d}\chi_L \;,\quad  
\chi_L(t)=\frac{1}{\sqrt{M}} \sum_k b_k^{({\rm R})} \;,
\end{equation}
and the equation identical to Eq.~(\ref{c1}) for the variables $b_k^{({\rm R})}$ of the right ring.

\subsection{Dependence on the system parameters}

\begin{figure}
\includegraphics[width=8.5cm,clip]{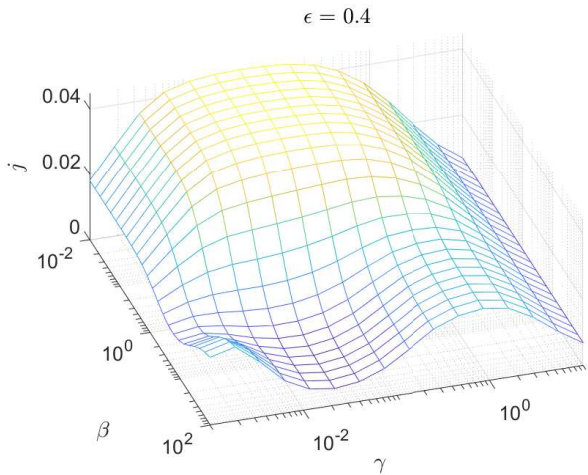}
\caption{The stationary current of non-interacting bosons in the chain connecting two rings with the mean particle density $\bar{n}_{\rm L}=1$ and $\bar{n}_{\rm R}=0.1$. The value of the coupling constant $\epsilon=0.4$.}
\label{fig2}
\end{figure}

Fig.~\ref{fig2} shows a typical dependence of the stationary current on the relaxation constant $\gamma$ and the inverse temperature $\beta$. As expected, the current vanishes for $\gamma\rightarrow0$ where $\bar{j}\sim\gamma$. Less expected is that the current also vanishes for  $\gamma\rightarrow\infty$. Formally, one proves this result by employing the Born and Markov approximations which are always justified in the above limit and which allow us to eliminate the rings. This reduces the system (\ref{a1}) to the standard open Bose-Hubbard model \cite{112} which is parametrised  by the effective relaxation constant 
\begin{equation}
\label{b6}
\tilde{\gamma}=\epsilon^2/\gamma \;.
\end{equation}
From Eq.~(\ref{b6}) we obtain that for  $\gamma\rightarrow\infty$ the current, which is now proportional to $\tilde{\gamma}$, decreases as $\bar{j}\sim 1/\gamma$. The details of this formal analysis are given in the Appendix. 

\begin{figure}[t]
\includegraphics[width=8.5cm,clip]{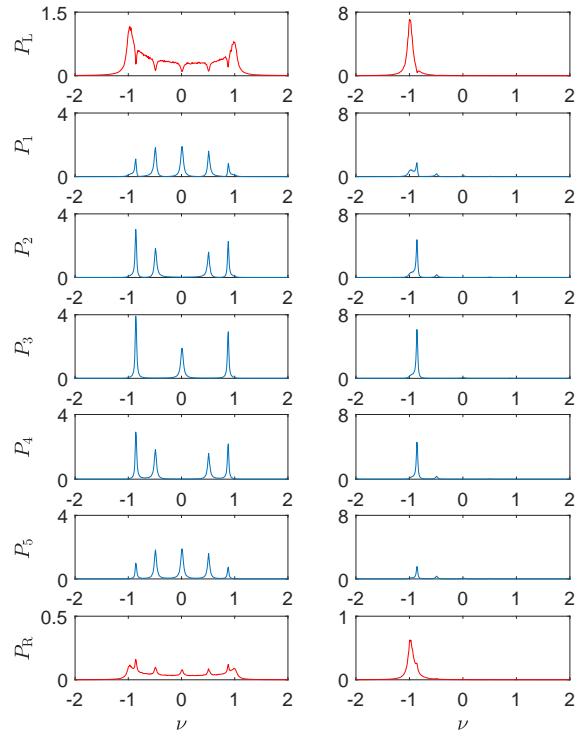}
\caption{The spectral densities $P(\nu)$ of the stochastic forces $\chi_1(t)$, upper panels, and $\chi_L(t)$, lower panels, for $\beta=0.1$, left column, and $\beta=10$, right column. The middle panels show the spectral densities of the local oscillators $a_\ell(t)$. The values of the relaxation and coupling constants are $\gamma=0.1$ and $\epsilon=0.4$.  The particle densities in the left and right reservoirs are $\bar{n}_{\rm L}=1$ and  $\bar{n}_{\rm R}=0.1$, respectively. Notice the different upper limits of the $y$-axis.}
\label{fig3}
\end{figure}

Next we address the dependence of the stationary current on the temperature. For moderate $\gamma$ one sees in Fig.~\ref{fig2} the pronounced step at $\beta\sim J$ where the current drops one order of magnitude. This step is caused by the boson condensation at low temperatures and the pseudo-classical approach provides the necessary details. The lower and upper panels in Fig.~\ref{fig3} show the spectral densities of the stochastic forces $\chi_1(t)$ and $\chi_L(t)$  while the middle panels are the spectral densities of the oscillators, i.e., the squared Fourier transform of $a_\ell(t)$. The left column in Fig.~\ref{fig3} refers to the case  $\beta=0.1$. One can easily identify in the figure the eigenfrequencies  $\omega_i$ and eigen-modes $X^{(i)}$ of the isolated chain
which are obtained by diagonalising the chain single-particle Hamiltonian $H_s$, 
\begin{equation}
\label{b7}
H_s X^{(i)}=  \omega_i X^{(i)}  \;.
\end{equation}
The positions of the peaks, which are well approximated by the Lorentzian of the width $\sim \epsilon^2$, coincides with $\omega_i$ while the peak heights are proportional to  $|X_\ell^{(i)}|^2$. Notice that for the currently considered $\beta$ the the broad-band stochastic force excited all eigen-modes of the chain.  The left column in Fig.~\ref{fig3} should be compared with the right column which refers to the case $\beta=10$. Here the narrow-band stochastic force is capable to excite only the lowest mode. Since the group velocity at the bottom of the conductance band tends to zero, we have much smaller current in the low-temperature limit. In the depicted numerical data one also sees the effect of the chain back action on the rings. This back action is obviously  smaller for smaller $\epsilon$. However, even for the considered $\epsilon=0.4$ it can be safely neglected. In the other words, in the analytical studies of the problem the spectral density of the stochastic force can be approximated by that shown in Fig.~\ref{fig1}, see Appendix.

Finally, let us discuss the dependence of the current on the particle density in the reservoirs. It follows from the general arguments that the stationary current across the chain is proportional to the difference $\bar{n}_{\rm L} -\bar{n}_{\rm R}$.  It is also well know that for larger particle density the condensation of bosons occurs at higher temperature. Thus, the proportional increase of the parameters $\bar{n}_{\rm L}$ and  $\bar{n}_{\rm R}$ will result in the proportional increase of the current. We check that for $\bar{n}_{\rm L}=2$ and  $\bar{n}_{\rm R}=0.2$ the resulting figure is almost indistinguishable from Fig.~\ref{fig2} after scaling  the vertical axis by the factor 2 and the proper (non-linear) scaling of the temperature axis.  
\begin{figure}[t]
\includegraphics[width=8.5cm,clip]{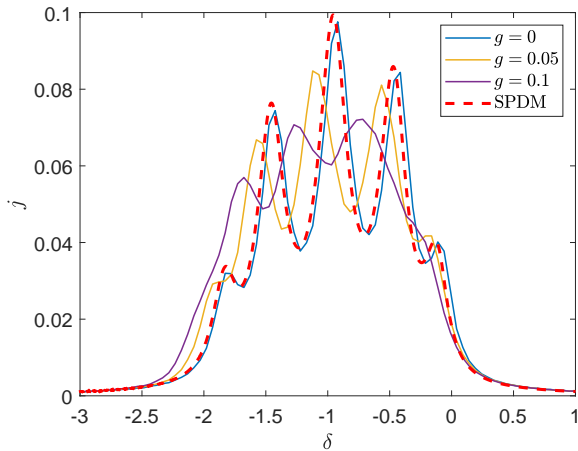}
\caption{Stationary current as the function of the gate voltage $\delta$ obtained by simulating the system dynamics over the time interval $1000T$ during which the gate voltage changes linearly in the interval $-3J\le\delta\le J$. (The other system parameters are $\beta=10$, $\gamma=0.2$, and $\epsilon=0.4$.) The dashed line is the SPDM solution, the solid lines are the result of the pseudo-classical approach (averaging over 6300 realizations) for $g=0,0.05,0.1$.}
\label{fig4}
\end{figure}

\subsection{Resonant transmission}

The resonant transmission of the fermions is a well studied phenomenon in the solid-state physics \cite{Datt95}. It occurs when the Fermi energy of the reservoirs coincides with an eigenenergy of the mesoscopic device. It is interesting to address the same phenomenon for bosons. Following  this goal we incorporate in all equations the gate voltage $\delta$ which determines the onsite energy in the Hamiltonian (\ref{a1}).

Similar to the case of fermions, there are necessary conditions for observing this effect, where the crucial  condition is that the distance between the energy levels of the mesoscopic device is smaller than the width of the Bose-Einstein distribution in the reservoirs. For the considered through the paper setup these conditions are satisfied, for example, for the parameters used in the right column in Fig.~\ref{fig3}. Thus,  we may expect a resonance-like behaviour of the stationary current under variation of the gate voltage $\delta$. Numerical simulations of the system dynamics fully confirms this expectation, see dashed line in Fig.~\ref{fig4}. Notice that positions of the resonant peak are slightly shifted from the expected $\delta=-(\omega_i+J)$. This shift increases with the increase of $\epsilon$ or decrease of $\gamma$ and is due to the back action of the chain on the rings. 

Next we study the effect of inter-particle interactions on the observed resonant transmission. To do this we use the pseudoclassical approach where the interparticle interactions are characterized by the macroscopic interaction constant $g=U\bar{n}$ where, as the parameter $\bar{n}$, we choose the mean particle density in the left reservoir. It is known that, the pseudo-classical approach is exact for $g=0$ and, if $g\ne0$, in the limit $\bar{n} \rightarrow\infty$ and  $U=g/\bar{n}\rightarrow 0$. Thus, for a fixed $U$ the method gives correct  results only up to  some critical $\bar{n}$. 

The blue solid line in Fig.~\ref{fig4} corresponds to $g=0$ ($U=0$) where the deviation from the dashed line is due to finite number of realisations of the stochastic force. This deviation indicates the statistical error for the chosen ensemble with 6300 realisations. It is seen in Fig.~\ref{fig4} that with increase of $g$ the resonant peaks shift to lower values of the gate voltage and, simultaneously,  the resonance pattern fades away. 

\section{Conclusions}
\label{sec4}

We analyzed the current of Bose particles across one-dimensional lattice connected at both sides to particle reservoirs. The lattice is modelled by the Bose-Hubbard chain and the reservoirs by bosons in the tight-binding rings which, if rings being disconnected from the lattice, relax with the rate $\gamma$ to the equilibrium state described by the Bose-Einstein distribution. For simplicity  we considered the case of equal reservoir temperatures. Then the stationary  current in the lattice is proportional to the difference in the mean particle densities of the reservoirs with the prefactor depending on the temperature $1/\beta$, relaxation rate $\gamma$, and the gate voltage $\delta$.  The central result of the work is that at low temperature the current as the function of the gate voltage can snow pronounced oscillations where it is significantly enhanced for the values of $\delta$ coinciding with the eigenenergies of the quantum particle in the  isolated lattice. Thus, similar to the  case of fermionic carries, we meet the phenomenon of the resonant transmission. We also quantify the role of interparticle interactions on the observed effect  and show that the resonance-like  pattern for the current gradually fades out with increasing the interaction constant. 

This work has been supported through Russian Science Foundation (RU), N19-12-00167.

\section{Appendix}

For non-interacting particles the derivation of the master equation for SPDM Eq.~(\ref{a9}) is similar to that for fermionic carries \cite{preprint}. Of course, one can obtain the master equation in the closed form only under the certain assumptions which are known as the Markov and Born approximations. The former approximation neglects the memory effect in the system dynamics, the latter neglects the back action of the chain on reservoirs. This means, in particular, that SPDM of the Bose particles in the reservoirs can be approximated by the thermal density matrix $\hat{\rho}_{\rm r}^{(0)}$ determined by the Bose-Einstein distribution Eq.~(\ref{a8}).
\begin{figure}[t]
\includegraphics[width=8.5cm,clip]{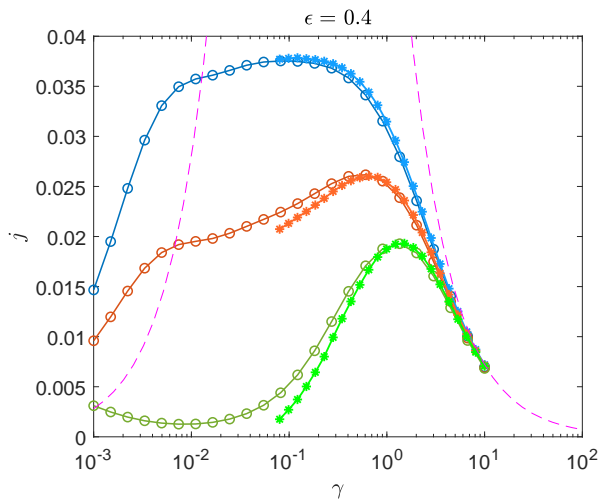}
\caption{Comparison of the results obtained on the basis of the Markovian, Eq.~ (\ref{d5}), and non-Markovian, Eq.~ (\ref{d1}), master equations with the exact results obtained on the basis of the original model.}
\label{fig5}
\end{figure}

According to Ref.~\cite{preprint} the master equation for the SPDM of the carries in the chain in the Born approximation reads
\begin{equation}
\label{d1}
\frac{\partial \hat{\rho}_{\rm s}}{\partial t}=-i[\widehat{H}_{\rm s},\hat{\rho}_{\rm s}]
+\epsilon^2\sum_{\ell=1,L}\left(\widehat{L}_{\ell}+ \widehat{L}_{\ell}^{\dagger}\right),
\end{equation}
where $\widehat{H}_{\rm s}$ is the chain single-particle Hamiltonian,
\begin{equation}
\label{d2}
\widehat{H}_{\rm s}= \delta \sum_{\ell=1}^{L}| \ell \rangle\langle \ell |
-\frac{J_{\rm s}}{2}\sum_{\ell=1}^{L-1}\left( |1\!+\!\ell\rangle\langle \ell|+\rm {h.c.}\right) \;,
\end{equation}
and the operators $\widehat{L}_{\ell}$ have the form
 \begin{equation}
 \label{d3}
\widehat{L}_\ell\!=\!\frac{| \ell \rangle\langle \ell |}{4} \!\int\limits_{-t}^{0}\!d\tau
e^{\frac{\gamma}{2}\tau}\!
\left[{\cal J}_{\rm F}(J_{\rm r}\tau)
\widehat{\mathbb I}_{\rm s}\!-\!{\cal J}_{0}(J_{\rm r}\tau)\hat{\rho}_{\rm s}(\tau\!+\!t) 
\!\right]\widehat{U}_{\rm s}(\tau) \;.
\end{equation} 
In Eq.~(\ref{d3})  ${\cal J}_0$ is the zeroth order Bessel function of the first kind, $\widehat{\mathbb I}_{\rm s}$ is the identity matrix of the size $L\times L$, $\widehat{U}_{\rm s}(\tau)$ denotes the evolution operator, $\widehat{U}_{\rm s}(\tau)=\exp(-i \widehat{H}_{\rm s} \tau)$, and
\begin{equation}
\label{d4}
{\cal J}_{\rm F}(J_{\rm r} t)=\frac{1}{2\pi}\int\limits_{-\pi}^{\pi}d\kappa\frac{e^{-iJ_{\rm r}\cos(\kappa)t}} 
{e^{-\beta[J_{\rm r}\cos(\kappa)+\mu]}-1} \;.
\end{equation}

If we now neglect the memory effects the depicted integrodifferential master equations transforms into the Markovin master equation. Formally this is done by using the general relation for any slowly varying function,
\begin{equation}
\int\limits_{0}^{t}d\tau
e^{\frac{\gamma}{2}\tau } {\cal A}(\tau+t) \approx \frac{2}{\gamma}{\cal A}(t) \;,
\end{equation}
which becomes exact in the limit $\gamma\rightarrow\infty$. This gives
\begin{equation}
\label{d5}
\frac{\partial \hat{\rho}_{\rm s}}{\partial t}\!=\!-i[\widehat{H}_{\rm s},
\hat{\rho}_{\rm s}]\!-\tilde{\gamma}\sum_{\ell=1,L}\!
\left(\frac{1}{2}\left\{|\ell \rangle\langle \ell|,\hat{\rho}_{\rm s} \right\}
\!-\!\bar{n}_{\ell}|\ell \rangle\langle \ell|\right)\!,
\end{equation}
where $\tilde{\gamma}=\epsilon^2/\gamma$.  Eq.~(\ref{d5}) is the SPDM master equation of the standard open BH model. It admits the analytical solution with the following result for the stationary current \cite{112}
\begin{equation}
\label{d6}
\bar{j}=J_{\rm s}\frac{J_{\rm s}\tilde{\gamma}}{J_{\rm s}^2 + \tilde{\gamma}^2}\frac{\bar{n}_{\rm L}-\bar{n}_{\rm R}}{2} \;.
\end{equation}

It is interesting to discuss the validity of the Born and Markov approximations with respect to the numerical data presented in Fig.~\ref{fig2}. The solid lines in Fig.~\ref{fig5} are the `cuts' of the latter figure for $\beta=0.1$ (high temperature), $\beta=1$ (moderate temperature), and $\beta=10$ (low temperature).  The dashed magenta line is the result obtained on the basis of the Markovian master equation (\ref{d5}). It is seen that the Markov approximation is justified only for the large $\gamma>5$. Unlike the Markovian master equation, non-Markovian master equation (\ref{d1}) is seen to be valid till $\gamma\approx0.1$. Thus it is capable, in particular, to describe the phenomenon of the resonant transmission considered in Sec.~\ref{sec3}.


\end{document}